\documentclass[prl,floatfix,twocolumn,showpacs,amsmath,amssymb]{revtex4}
\usepackage{graphicx}
\usepackage{dcolumn}
\usepackage{bm}

\begin{document}
\title{Superfluid to Mott-insulator transition in Bose-Hubbard models}

\author{Manuela Capello,$^{1}$ Federico Becca,$^{2,3}$ Michele Fabrizio,$^{2,3,4}$
Sandro Sorella,$^{2,3}$}
\affiliation{
$^{1}$ Laboratoire de Physique Th\'eorique, Universit\'e Paul Sabatier, CNRS, 31400 Toulouse, France\\
$^{2}$  International School for Advanced Studies (SISSA),  
I-34014 Trieste, Italy \\
$^{3}$ CNR-INFM-Democritos National Simulation Centre, Trieste, Italy. \\
$^{4}$ International Centre for Theoretical Physics (ICTP), P.O. Box 586, I-34014 Trieste, Italy
}

\date{\today}

\begin{abstract}
We study the superfluid-insulator transition in Bose-Hubbard models in one-, 
two-, and three-dimensional cubic lattices by means of a recently proposed variational 
wave function. In one dimension, the variational results agree with the expected 
Berezinskii-Kosterlitz-Thouless scenario of the interaction-driven Mott transition. 
In two and three dimensions, we find evidences that, across the transition,  
most of the spectral weight is concentrated at high energies, suggestive of pre-formed 
Mott-Hubbard side-bands. This result is compatible with the experimental data by 
Stoferle {\it et al.} [Phys. Rev. Lett. {\bf 92}, 130403 (2004)]. 

\end{abstract}


\maketitle

Recent experiments on cold atoms trapped in optical lattices demonstrated that the Mott 
transition (MIT), originally introduced in electronic systems,~\cite{mott} can 
be experimentally realized also in bosonic systems,~\cite{greiner} where the MIT is 
actually a superfluid-insulator transition. Recent experiments by Stoferle 
{\it et al.}~\cite{stoferle} have shown that a considerable amount of spectral weight is 
concentrated at high energy even within the superfluid phase. More specifically, the data 
suggest that, especially in three-dimensions, a Mott-Hubbard gap of order $U$ develops already 
on the superfluid side of the MIT, akin to what is predicted to occur in 
electronic systems.~\cite{DMFT} Although these evidences are not incompatible with the 
accepted theory of the critical behavior across the superfluid-to-insulator 
transition,~\cite{fisher} they clearly demand for a more detailed comprehension that 
must include also high-energy excitations. There have been already several theoretical 
attempts, mainly based on suitable extensions of mean-field theory, to uncover the whole 
dynamical behavior across the MIT.~\cite{blatter,clark}
These calculations predict in the most general cases the existence of high energy modes 
even in the superfluid phase that might explain the experimental data. However, when the 
transition is approached at fixed integer filling by tuning for instance the interaction 
strength, these theories also predict that all modes soften at the transition. 
In this work, we intend to address this question by an alternative approach based on 
a variational wave function that has been recently proposed in the context of the 
electronic MIT.~\cite{capello,capello2,capello3} 
The accuracy of the wave function is checked by comparison with Green's Function 
Monte Carlo (GFMC) simulations, that allow us to obtain numerically exact results by a 
stochastic sampling of the ground-state wave function.~\cite{calandra} In contrast with 
the aforementioned mean-field theories, we find that, at fixed density, the MIT
is accompanied by a gradual transfer of spectral weight from the low-energy sound mode 
towards high energies, so that, when the Mott insulating phase is established, 
most of the spectral weight is already concentrated at high energy.
In addition, our analysis uncovers features of variational wave functions able to 
describe a Mott transition that are novel and might be common to bosonic as well as 
fermionic systems.

Bosons in optical lattices can be modeled by the Hubbard 
Hamiltonian:~\cite{jaksch,fisher,roth,damski}
\begin{equation}\label{hambose}
{\cal H} = - \sum_{ij,\,\sigma}\, 
\bigg(t_{ij}\,a^\dagger_{i\sigma} a^{\phantom{\dagger}}_{j\sigma} + H.c.\bigg)
+ \frac{U}{2} \sum_i \, n_i\, (n_i-1),
\end{equation}
where $a^\dagger_{i\sigma}$ ($a_{i\sigma}$) creates (annihilates) a particle at site $i$ 
with integer spin $\sigma=-S,\dots,S$, and $n_i=\sum_\sigma\,a^\dagger_{i\sigma} 
a^{\phantom{\dagger}}_{i\sigma}$.  
The Hubbard model~(\ref{hambose}) at integer filling has generally two different 
phases: one superfluid, for $U<U_c$, and the other insulating above $U_c$.  
In this work, we shall focus on the spinless case and
attempt to describe the MIT by means of a variational wave function. In spite of the 
fact that the variational approach is a simple and well established technique, 
its application to the MIT turns out to be extremely difficult. For instance, the 
celebrated Gutzwiller wave function is not appropriate to describe the MIT, as it leads to  
an unrealistic insulator with no density fluctuations.~\cite{kotliar,krauth2} 

\begin{figure}
\includegraphics[width=0.40\textwidth]{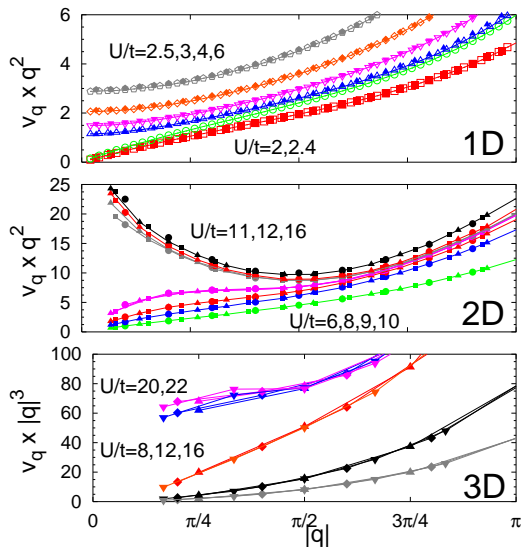}
\caption{\label{fig:jas}
Variational results for the Jastrow potential 
$v_q$ multiplied by $q^2$ in 1D and 2D and by $|q|^3$ in 3D for increasing values of $U/t$
(from bottom to top). Upper panel: 1D case for $60$ and $100$ sites. 
Middle panel: 2D case for $20 \times 20$, $26 \times 26$, and $30 \times 30$ clusters
(along the $(1,0)$ direction). Lower panel: 3D case for $8 \times 8 \times 8$, 
$10 \times 10 \times 10$, and $12 \times 12 \times 12$ clusters 
(along the $(1,0,0)$ direction).}
\end{figure}

Recently, an extension of the Gutzwiller wave function has been proposed~\cite{capello},  
that proved to be very accurate to 
describe an electronic MIT in 1D.~\cite{capello,capello3}  
Here we apply the same variational approach to the 
$S=0$ Bose-Hubbard model~(\ref{hambose}) with nearest-neighbor hopping $t/2$ in a 
one-dimensional chain (1D), a two-dimensional (2D) square lattice and a three-dimensional 
(3D) cubic lattice with $L$ sites and periodic boundary conditions. We consider the following 
ansatz for the variational wave function
\begin{equation}\label{wavefunction}
|\Psi \rangle = \exp \left ( -\frac{1}{2} \sum_{i,j} v_{i,j} n_i n_j 
+g_{MB} \sum_i \xi_i \right ) |\Phi_0 \rangle,
\end{equation}
where $|\Phi_0 \rangle$ is the non-interacting fully-condensed wave function, 
i.e. $|\Phi_0 \rangle = (b_{k=0}^\dagger)^N|0 \rangle$, being $b^\dagger_k$ the creation 
operator at momentum $k$ and $N=L$ the number of particles. The components of the Jastrow 
potential, $v_{i,j}= v(|R_i-R_j|)$, are independently optimized by a 
Variational Monte Carlo (VMC) minimization of the total energy.~\cite{sorella} In the following, 
we will denote by $\rho_q$ and $v_q$ the Fourier transforms of the boson-density $n_i$ and of the 
Jastrow parameters $v_{i,j}$, respectively. 
Finally, $g_{MB}$ is a variational parameter related to the many-body 
operator $\xi_i=h_i \prod_{\delta} (1-d_{i+\delta})+d_i \prod_{\delta} (1-h_{i+\delta})$,
where $h_i=1$ ($d_i=1$) if the site $i$ is empty (doubly occupied) and $0$
otherwise, and $\delta$ is the vector that connects nearest-neighbor sites.~\cite{kaplan} 
This term is kept just to improve the variational accuracy (mainly in 2D and 3D) but does 
not introduce important correlation effects, that are instead contained only in the 
{\it long-range} tail of the two-body Jastrow potential $v_{i,j}$. 
Both the Jastrow factor and the many-body operator $\xi$ commute with the particle number, hence 
(\ref{wavefunction}) belongs to the Fock space with $N=L$ bosons. 

\begin{figure}
\includegraphics[width=0.40\textwidth]{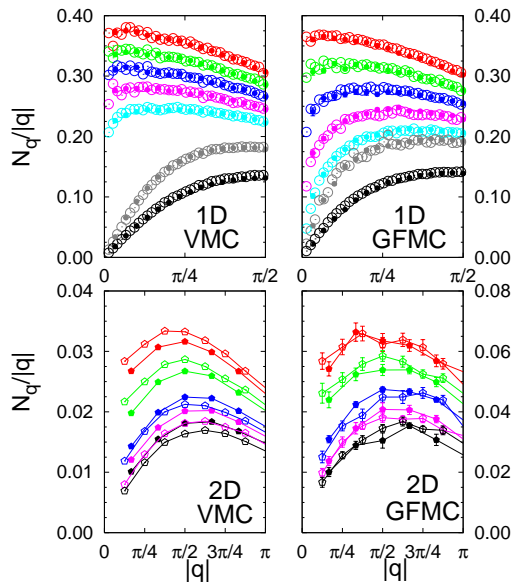}
\caption{\label{fig:nq1d2d}
Density structure factor $N_q$ divided by $|q|$ calculated with variational
Monte Carlo (left panels) and GFMC (right panels) in 1D (upper panels) and 2D (lower panels)
In 1D, $L=60$ (full symbols) and $L=150$ (empty symbols) and $U/t=1.6$, $1.8$, $2$, $2.2$, 
$2.4$, $2.5$ and $3$. In 2D, $L=12 \times 12$ (full symbols) and $16 \times 16$ 
(empty symbols) for the variational calculation ($U/t=10$, $10.2$, $10.4$, $10.6$, and $10.8$)
and for the GFMC calculation ($U/t=8$, $8.2$, $8.4$, $8.6$, and $8.8$). All cases are shown 
from top to bottom for increasing values of $U/t$.}
\end{figure}

Although our ultimate scope is to uncover some dynamical properties across the MIT, 
we think it is worth discussing in some detail the role of the Jastrow factor  
in~(\ref{wavefunction}), which is the novel ingredient with respect to the conventional 
Gutzwiller wave function. In the superfluid phase, a long-range 
Jastrow potential is surely needed to restore the correct small-$q$ behavior of the static 
density structure factor, i.e., 
$N_q=\langle \Psi | \rho_{-q} \rho_q |\Psi \rangle / \langle \Psi|\Psi \rangle \sim |q|$. 
Indeed, since at least at weak-coupling, the expression
\begin{equation}\label{Reatto&Chester}
N_q = \frac{\displaystyle N^0_q}{\displaystyle 1 + \gamma \,v_q\,N^0_q}
\end{equation}
holds with $\gamma=2$,~\cite{R&C} and because the non-interacting 
$N_q^0=\langle \Phi_0 | \rho_{-q} \rho_q |\Phi_0 \rangle \sim {\rm const}$, it follows that 
$v_q\sim 1/|q|$. Assuming that the expression~(\ref{Reatto&Chester}) remains valid 
for $|q| \to 0$ even in the Mott insulating phase, one might be 
tempted to believe that $v_q\sim 1/q^2$ is necessary and sufficient 
to recover in any dimension the appropriate $N_q \sim q^2$ insulating behavior, 
consequence of exponentially decaying correlation functions. 
However, one easily realizes that, were this conclusion correct, the 
variational wave function~(\ref{wavefunction}) could not describe 
any bosonic insulator in 3D, since $v_q \sim 1/q^2$ is not sufficient to empty the 
condensate fraction.~\cite{reatto} Indeed, as shown below, the optimized variational 
wave function has a more diverging $v_q\sim 1/|q|^3$ in the 3D Mott insulator, although 
$N_q\sim q^2$, implying that the formula~(\ref{Reatto&Chester}) does not generally hold.

In Fig.~\ref{fig:jas} we draw the optimized Jastrow potential $v_q$. For any 
dimension, the MIT is clearly signaled by the sudden change in the small-$q$ 
behavior of $v_q$. On the one hand, the superfluid phase is always described by 
$v_q \sim \alpha/|q|$, with $\alpha$ increasing with $U$. On the other hand, 
the Mott insulator has a much more diverging $v_q$. In 1D we recover the 
$v_q \sim 1/q^2$ behavior, like in the fermionic case.~\cite{capello}
In 2D, the leading behavior of the Jastrow potential across the transition is less 
clearcut than in 1D. Indeed, we cannot establish whether, on the insulating side, 
the leading behavior is given by $v_q\sim \beta_{2D}/q^2$ with $\beta_{2D}$ large but 
finite, or logarithmic corrections have to be considered, i.e., $v_q \sim \ln(1/|q|)/q^2$.
Finally, in 3D a more diverging $v_q\sim 1/|q|^3$ is stabilized in the insulating regime. 
Therefore, in all cases the Jastrow potential is able to empty the condensate.~\cite{notenk}
We note that, within this approach, the MIT shows up in the wavefunction in the form of a 
binding-unbinding transition of opposite-charged particles 
(empty and doubly occupied sites).~\cite{capello2} 

In order to check the validity of our approach, we compare the VMC results of the
density structure factor $N_q$ with the numerically exact ones obtained by GFMC. At small 
$q$'s we can generally write $N_q \sim \gamma_1 |q| + \gamma_2 q^2$. In the 
superfluid phase, $\gamma_1 \ne 0$ while, in the Mott insulator, $\gamma_1=0$ and 
$\gamma_2 \not = 0$, see Fig.~\ref{fig:nq1d2d}. 
In 1D, we have evidence that $\gamma_1$ has a very
sharp crossover from a finite value to zero across the MIT, suggestive of 
a true jump in the thermodynamic limit. Moreover, our numerical results 
indicate that $\gamma_2$ diverges as the MIT is approached from the insulating side 
(this is particularly evident from the GFMC results). Within 
the variational approach, this behavior follows from $v_q \sim \beta_{1D}/q^2$ in the 
insulating phase with $\beta_{1D} \to 0$ at the transition. 
In conclusion, the 1D MIT can be located at $U_c/t \simeq 2.45 \pm 0.05$ in the VMC, whereas 
GFMC gives $U_c/t \simeq 2.1 \pm 0.1$ (in agreement with previous 
calculations of Ref.~\cite{batrouni,kuhner}), showing that the wave function~(\ref{wavefunction}) 
is not only qualitatively but also quantitatively correct. 

\begin{figure}
\includegraphics[width=0.40\textwidth]{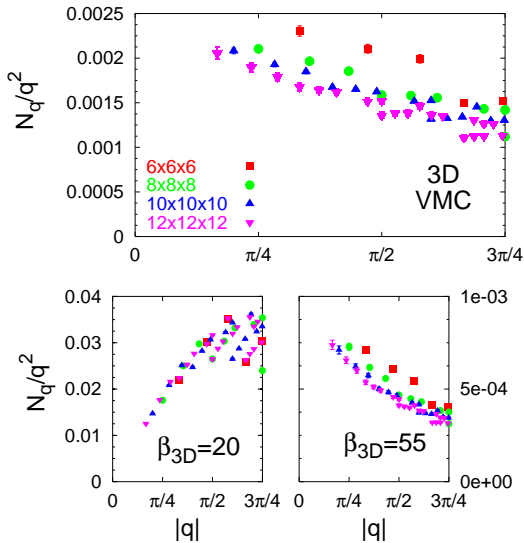}
\caption{\label{fig:nq3d}
Upper panel: Density structure factor $N_q$ calculated by the variational Monte Carlo 
for 3D and $U/t=20$. Lower panels: $N_q$ for non-optimized wave functions with 
$v_q \sim \beta_{3D}/|q|^3$ for two values of $\beta_{3D}$.}
\end{figure}

\begin{figure}
\includegraphics[width=0.40\textwidth]{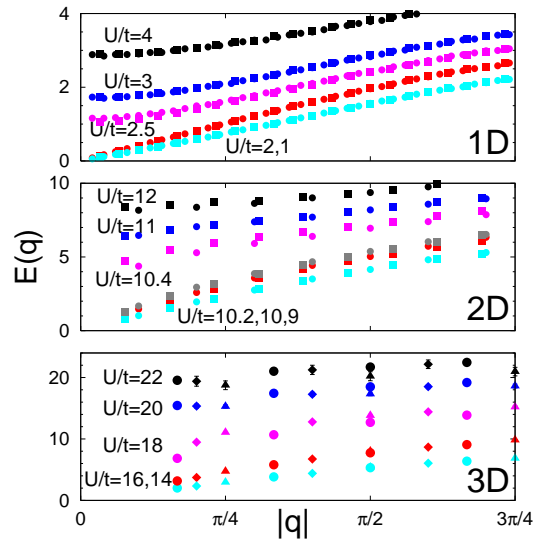}
\caption{\label{fig:spectrum}
Variational results for the average energy of density excitations $E(q)$ 
of Eq.~(\ref{Delta_q}) in 1D, 2D, and 3D.}
\end{figure}

The density structure factor $N_q$ displays quite distinct long-wavelength 
behaviors for weak and strong interaction also in 2D, see Fig.~\ref{fig:nq1d2d}. 
The VMC structure factor goes like $N_q \sim \gamma_1 |q| + \gamma_2 q^2$, for $U/t \lesssim 10.3$, 
while above we find $N_q \sim \gamma_2 q^2$. The critical value of the on-site interaction is
slightly different from the GFMC one, $U_c/t \simeq 8.5$, which agrees with 
Ref.~\cite{krauth}.
In spite of slightly different values of $U_c$, the qualitative behavior across the MIT is 
similar both in VMC and in GFMC. Differently from 1D, 
approaching the MIT in 2D, $\gamma_1 \to 0$ while $\gamma_2\not = 0$ is smooth across the transition.

Even more interesting is the 3D case. Here, the GFMC is severely limited by small sizes and,
therefore, we just discuss the variational results. As we mentioned, the optimal Jastrow 
potential turns from $v_q\sim \alpha/|q|$ in the superfluid phase, into 
$v_q\sim \beta_{3D}/|q|^3$ in the Mott insulator (see Fig.~\ref{fig:jas}). 
The sudden change of behavior allows to locate the transition around $U_c/t \simeq 18$, 
which is close to the critical value of recent Monte Carlo simulations in 
3D.~\cite{prokofev} Even though $v_q\sim 1/|q|^3$, the structure factor in the Mott 
insulator has the correct behavior $N_q\sim q^2$. In turns this implies that 
Eq.~(\ref{Reatto&Chester}) does not hold, not even for $|q|\to 0$, which is 
quite unexpected. In order to prove more firmly that $v_q\sim \beta_{3D}/|q|^3$ can indeed 
lead to $N_q\sim q^2$, we have calculated the latter with a non-optimized wave function of the 
form~(\ref{wavefunction}) with $v_q\sim \beta_{3D}/|q|^3$ and for different values of 
$\beta_{3D}$.
As shown in Fig.~\ref{fig:nq3d}, for small $\beta_{3D}$ we have $N_q\sim |q|^3$, implying 
that Eq.~(\ref{Reatto&Chester}) is qualitatively correct. However, above a critical 
$\beta^*_{3D}$, the behavior turns into $N_q \sim q^2$, signaling a remarkable breakdown 
of Eq.~(\ref{Reatto&Chester}).
The optimal value of $\beta_{3D}$ that we get variationally at the MIT is larger than 
$\beta^*_{3D}$, confirming our variational finding $N_q \sim q^2$. We note that the 
change of behavior as a function of $\beta_{3D}$ is consistent with the binding-unbinding 
phase transition recently uncovered in a classical 3D gas with
$1/|q|^3$-potential.~\cite{3dcgm} 

Let us now come back to our original motivation concerning the dynamical properties across 
the superfluid-insulator transition. Although our variational wave function is 
meant only to describe the ground state, it can also provide important insights into the 
structure of the excitation spectrum. One can easily prove that the following expression holds 
in model (\ref{hambose}): 
\begin{equation}\label{Delta_q}
E(q) = 2 \sum_{i=1}^{D} \sin^2\left(\frac{q_i}{2}\right)\; 
\frac{\displaystyle \langle -T \rangle}{\displaystyle D\,N_q} = 
\frac{\displaystyle \int d\omega\; \omega\; S(q,\omega)}{\displaystyle
\int d\omega\;  S(q,\omega)},
\end{equation}
where $\langle T \rangle$ is the average value of the hopping, the sum is over the
spatial directions, $D=1,2,3$ is the 
space dimensionality, and $S(q,\omega)$ the dynamical structure factor. 
$E(q)$ is the first moment of $S(q,\omega)$ and can be regarded as the {\it average energy} 
of density excitations, which is therefore directly accessible through our variational 
calculation. In the superfluid phase $E(q)\propto q$ at small $q$, while $E(q)$ develops a 
finite gap, i.e., $E(0)\not =0$, in the Mott insulator, which is an upper bound of the actual 
Mott-Hubbard gap. In 1D this gap seems to vanish at the MIT, in agreement 
with the Berezinskii-Kosterlitz-Thouless scenario, see Fig.~\ref{fig:spectrum}. 
On the contrary, both in 2D and 3D, we find that $E(0)$ is finite and of order $U$ 
everywhere in the Mott insulating side, even right at the MIT, see Fig.~\ref{fig:spectrum}. 
This implies that high-energy excitations exist in the Mott insulator and carry most of 
the spectral weight. Also interesting is the behavior of the linear slope of $E(q)$ within 
the superfluid phase as the MIT is approached. We recall that, assuming for the structure 
factor the small-$q$ expression $N_q = \gamma_1\, |q| + \gamma_2\,q^2 + O(q^3)$, $\gamma_1$
is finite at the MIT in 1D, while it vanishes in 2D and 3D. Moreover, as the MIT is approached, 
$\gamma_2$ diverges in 1D but stays finite in 2D and 3D. 
Since the hopping energy is finite and continuous across the 
transition, it follows that in 2D and 3D the linear slope of $E(q)$ should diverge at the MIT, 
although our numerical evidence is more clearcut in 3D than in 2D. 
Excluding the possibility that the sound velocity diverges at the MIT, we must conclude 
that the spectral weight is gradually transferred from the sound mode to high-energy 
excitations that exist already in the superfluid phase and are smooth across the MIT, 
suggestive of pre-formed Mott-Hubbard side bands. These results are actually consistent 
with the experimental data of Ref.~\cite{stoferle}.   

In conclusion we have demonstrated that a long-range Jastrow potential does allow for a 
faithful variational description of a Mott transition in the bosonic Hubbard model.  
The average energy of the charge-density excitations, that is accessible by our calculation,
suggests in two and three dimensions that pre-formed Hubbard side-bands coexist with 
sound modes in the superfluid phase near the Mott transition and carry most of the spectral 
weight in the insulator. This is an interesting and also surprising result, that bears a 
lot of similarities with the MIT in electronic systems,~\cite{DMFT} but is not accounted for 
by most accepted theories of the superfluid to Mott insulator transition in bosonic systems. 
In analogy with the bosonic example, we expect that a singular
Jastrow potential $v_q\sim 1/|q|^3$ is necessary to describe the 3D Mott transition 
in fermionic models, too, all the more reason when realistic Coulomb interaction is taken 
into account, in which case a Jastrow potential $1/q^2$ is necessary already in the metal to 
reproduce the proper long-wavelength behavior.  
 
We acknowledge useful discussions with D. Poilblanc and T. Senthil.
This work has been partially supported by CNR-INFM and
COFIN 2004 and 2005.


\begin{thebibliography}{99}

\bibitem{mott} N.F. Mott, {\it Metal Insulator Transition}
   (Taylor and Francis, London, 1990).
\bibitem{greiner} M. Greiner {\it et al.}, Nature (London) {\bf 415}, 39 (2002).
\bibitem{stoferle} T. Stoferle {\it et al.}, \prl {\bf 92}, 130403 (2004).
\bibitem{DMFT} A. Georges {\it et al.}, \rmp {\bf 68}, 13 (1996).
\bibitem{fisher} M.P.A. Fisher {\it et al.}, \prb {\bf 40}, 546 (1989).
\bibitem{blatter} S.D. Huber {\it et al.}, \prb {\bf 75}, 085106 (2007).
\bibitem{clark} S.R. Clark and D. Jaksch, \pra {\bf 70}, 063612 (2004).
\bibitem{capello} M. Capello {\it et al.}, \prl {\bf 94}, 026406 (2005).
\bibitem{capello2} M. Capello {\it et al.}, \prb {\bf 73}, 245116 (2006).
\bibitem{capello3} M. Capello {\it et al.}, \prb {\bf 72}, 085121 (2005).
\bibitem{calandra} M. Calandra Buonaura and S. Sorella, \prb {\bf 57}, 11446 (1998).
\bibitem{jaksch} D. Jaksch {\it et al.}, \prl {\bf 81}, 3108 (1998).
\bibitem{roth} R. Roth and K. Burnett, J. Phys. B {\bf 37}, 3893 (2004).
\bibitem{damski} B. Damski and J. Zakrzewski, \pra {\bf 74}, 063609 (2006).
\bibitem{kotliar} D.S. Rokhsar and B.G. Kotliar, \prb {\bf 44}, 10328 (1991).
\bibitem{krauth2} W. Krauth, M. Caffarel, and J.P. Bouchaud, \prb {\bf 45}, 3137 (1992).  
\bibitem{sorella} S. Sorella, \prb {\bf 71}, 241103 (2005).
\bibitem{kaplan} T.A. Kaplan, P. Horsch, and P. Fulde, \prl {\bf 49}, 889 (1982). 
\bibitem{R&C} T. Gaskell, Proc. Phys. Soc. {\bf 77}, 1182 (1961); L. Reatto and 
   G.V. Chester, Phys. Rev. {\bf 155}, 88 (1967).
\bibitem{reatto} L. Reatto, Phys. Rev. {\bf 183}, 334 (1969).
\bibitem{notenk} In 1D, even $v_q \sim 1/|q|$ is sufficient to completely suppress the
   condensate fraction, as expected from Ref.~\cite{reatto}.
\bibitem{batrouni} G.G. Batrouni, R.T. Scalettar, and G.T. Zimanyi, \prl {\bf 65}, 
   1765 (1990).
\bibitem{kuhner} T.D. Kuhner and H. Monien, \prb {\bf 58}, R14741 (1998).
\bibitem{krauth} W. Krauth and N. Trivedi, Europhys. Lett. {\bf 14}, 627 (1991).
\bibitem{prokofev} B. Capogrosso-Sansone, N.V. Prokof'ev, and B.V. Svistunov, 
   \prb {\bf 75}, 134302 (2007).
\bibitem{3dcgm} S. Kragset, A. Sudbo, and F.S. Nogueira, \prl {\bf 92}, 186403 (2004).

\end{thebibliography}
\end{document}